\documentclass[aps,amsfonts,nofootinbib,twocolumn,showpacs]{revtex4}
\usepackage{graphicx}
\usepackage{amsmath}
\def\be{\begin{equation}}
\def\ee{\end{equation}}
\def\bea{\begin{eqnarray}}
\def\eea{\end{eqnarray}}
\def\orc{\Omega_{r_c}}
\def\om{\Omega_{\text{m}}}

\begin{document}

\title{Observational constraints on self-accelerating cosmology}

\author{Roy Maartens and Elisabetta Majerotto}

\affiliation{\vspace*{0.2cm} Institute of Cosmology \&
Gravitation, University of Portsmouth, Portsmouth~PO1~2EG, UK
\vspace*{0.2cm}}

\date{\today}

\begin{abstract}

The DGP brane-world model provides a simple alternative to the
standard LCDM cosmology, with the same number of parameters. There
is no dark energy -- the late universe self-accelerates due to an
infrared modification of gravity. We compute the joint constraints
on the DGP model from supernovae, the cosmic microwave background
shift parameter, and the baryon oscillation peak in the SDSS
luminous red galaxy sample. Flat DGP models are within the joint
2~sigma contour, but the LCDM model provides a significantly
better fit to the data. These tests are based on the background
dynamics of the DGP model, and we comment on further tests that
involve structure formation.

\end{abstract}

 \maketitle

\section*{1. THE DARK ENERGY PROBLEM}

The acceleration of the late-time universe, as implied by
observations of supernovae, cosmic microwave background
anisotropies and the large-scale structure, poses one of the
deepest theoretical problems facing
cosmology~\cite{Copeland:2006wr}. Within the framework of general
relativity, the acceleration originates from a dark energy field
(or effective dark energy) with negative pressure ($w \equiv
p/\rho<-{1\over3}$), such as vacuum energy ($w=-1$) or a
slow-rolling scalar field (``quintessence", $w>-1$). So far, none
of the available models has a natural explanation.

For the simplest option of vacuum energy, i.e., the LCDM model,
the incredibly small value of the cosmological constant
 \bea
&& \rho_{\Lambda,\text{obs}}={\Lambda \over 8\pi G}\sim
H_0^2M_P^2\sim (10^{-3}~\mbox{eV})^4 \,, \\ &&
\rho_{\Lambda,\text{theory}}\sim
 M_{\text{fundamental}}^4 > (1~\mbox{TeV})^4 \gg
\rho_{\Lambda,\text{obs}}\,,
 \eea
cannot be explained by current particle physics. In addition, the
value needs to be fine-tuned,
\begin{equation}
\Omega_{\Lambda}\sim \om\,,
\end{equation}
which also has no natural explanation. Quintessence models attempt
to address the fine-tuning problem, but do not produce a natural
solution -- and also cannot address the problem of how $\Lambda$
is set exactly to 0.

Alternatively, it is possible that there is no dark energy, but
instead an infrared modification of general relativity, i.e., on
very large scales, $r\gtrsim H_0^{-1}$, that accounts for
late-time acceleration. (Note that this does not remove the
problem of explaining why the vacuum energy does not gravitate.)
Schematically, we are modifying the geometric side of the field
equations,
 \be
G_{\mu\nu}+G^{\rm dark}_{\mu\nu} = 8\pi G T_{\mu\nu}\,,
\label{mod}
 \ee
rather than the matter side,
 \be
G_{\mu\nu} = 8\pi G \left(T_{\mu\nu}+ T^{\rm dark}_{\mu\nu}
\right)\,,
 \ee
as in general relativity. It is important that the modification is
covariant and incorporates deviations from homogeneity and
isotropy, so that one can compute not only the background
dynamics, but also the perturbations.

\section*{2. DGP MODIFIED GRAVITY}

One of the simplest covariant modified-gravity models is based on
the Dvali-Gabadadze-Porrati (DGP) brane-world model, as
generalized to cosmology by Deffayet~\cite{Dvali:2000rv}. (It is
worth noting that the original DGP model with a Minkowski brane
was not introduced to explain acceleration -- the generalization
to a Friedman brane was subsequently found to be
self-accelerating.) In this model, gravity leaks off the
4-dimensional brane universe into the 5-dimensional bulk spacetime
at large scales. At small scales, gravity is effectively bound to
the brane and 4D gravity is recovered to a good approximation, via
the lightest modes of the 5D graviton -- effectively via an
ultralight metastable graviton in the 4D universe. The transition
from 4D to 5D behaviour is governed by a crossover scale $r_c$;
the weak-field gravitational potential behaves as $r^{-1}$ for $
r\ll r_c$ and as $r^{-2}$ for $ r\gg r_c$.

The energy conservation equation remains the same as in general
relativity, but the Friedman equation is modified:
\begin{eqnarray}
&& \dot\rho+3H(\rho+p)=0\,,\label{ec} \\ && H^2+{K \over a^2}-{1
\over r_c}\sqrt{H^2+{K \over a^2}} = {8\pi G \over 3}\rho\,.
\label{f}
\end{eqnarray}
These equations imply (for the CDM case $p=0$)
 \be
\dot H- {K \over a^2}=-4\pi G\rho\left[ 1 + {1\over \sqrt{1+ 32\pi
G r_c^2\rho/3}} \right]. \label{r}
 \ee
Equation~(\ref{f}) shows that at early times, when $H^2 +K/a^2 \gg
r_c^{-2}$, the general relativistic Friedman equation is
recovered. By contrast, at late times in a CDM universe, with
$\rho\propto a^{-3}\to0$, we have
\begin{equation}
H\to H_\infty= {1\over r_c}\,.
\end{equation}
Gravity leakage at late times initiates acceleration -- not due to
any negative pressure field, but due to the weakening of gravity
on the brane. Since $H_0>H_\infty$, in order to achieve
self-acceleration at late times, we require
 \be
r_c\gtrsim H_0^{-1}\,,
 \ee
and this is confirmed by fitting observations, as discussed below.

\begin{figure}
\begin{center}
\includegraphics[width=7.5cm]{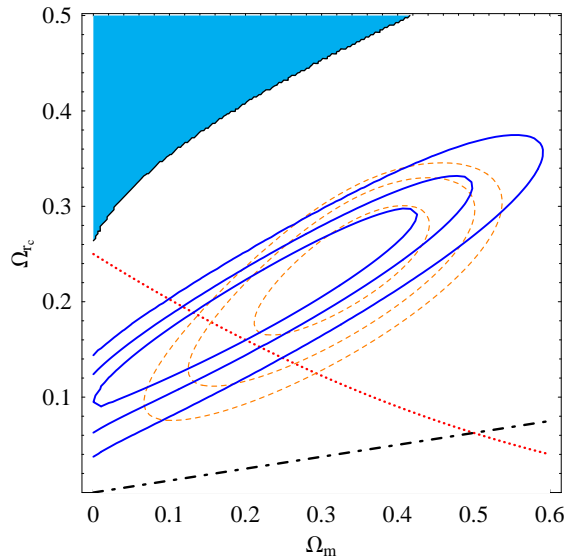}
\caption{The parameter space for DGP models. The dotted (red)
curve is the flat case, with closed models above and open models
below; the dot-dashed (black) curve demarcates models that are
currently accelerating (above) from those decelerating (below);
the shaded (blue) region contains expanding models that bounce in
the past, i.e. do not have a big bang; the likelihood contours are
fits to the SNe Gold data [dashed (brown)] and Legacy data [solid
(blue)]. }\label{plane}
\end{center}
\end{figure}

In dimensionless form, the modified Friedman equation~(\ref{f}) is
 \bea
{H(z)^2 \over H_0^2}&=&\left[ \sqrt{\om (1+z)^3+ \orc}+\sqrt{\orc}
\right]^2 \nonumber \\&&~~{}+\Omega_K(1+z)^2\,, \label{fz}
 \eea
where
 \bea
 \Omega_K &=& 1-\om-2\sqrt{\orc}\left(\!\sqrt{\orc}+\sqrt{\orc+\om}
\right)\!\!, \label{ok}\\
\orc &=& {1 \over 4 H_0^2 r_c^2}\,.
 \eea
From Eq.~(\ref{r}), the dimensionless acceleration is
 \bea
{1\over H_0^2}{\ddot a \over a}&=&\left( \sqrt{\om (1+z)^3+
\orc}+\sqrt{\orc}\right)\Biggl[ \sqrt{\orc} \nonumber\\ &&~{}+
{2\orc-\om(1+z)^3 \over 2\sqrt{\om (1+z)^3+ \orc} }\Biggr],
 \eea
so that the redshift when acceleration starts is given by
 \be
1+z_a=2\left( {\orc \over \om} \right)^{1/3}. \label{za}
 \ee
The $(\orc,\om)$ plane is illustrated in Fig.~\ref{plane}. This
shows the flat model curve [from Eq.~(\ref{ok})],
 \be
\orc ={1 \over 4}(1-\om)^2,
 \ee
and the present-time zero-acceleration curve [from
Eq.~(\ref{za})],
 \be
\orc={\om \over 8}\,.
 \ee
By checking for real roots $z>0$ of the equation $H(z)=0$, given
by Eq.~(\ref{fz}) with radiation included, we also find the region
of the plane containing bouncing models, i.e., models that are
currently expanding but were collapsing in the past.

The modified Friedman equation in DGP may reinterpreted from a
standard viewpoint. We define the effective dark energy density
$\rho_{\text{eff}}\equiv 3H/8\pi Gr_c$. Then the effective dark
energy equation of state $w_{\text{eff}}\equiv
p_{\text{eff}}/\rho_{\text{eff}}$ is given by
$\dot{\rho}_{\text{eff}} +3H(1+w_{\text{eff}})
\rho_{\text{eff}}=0$. Thus $\rho_{\text{eff}}$ and
$w_{\text{eff}}$ give a standard general relativistic
interpretation of DGP expansion history, i.e., they describe the
equivalent general relativity dark energy model. For the flat
case, $\Omega_K=0$, we find
 \be \label{wx}
w_{\text{eff}}(z)={\om-1-\sqrt{(1-\om)^2+4\om(1+z)^3} \over
2\sqrt{(1-\om)^2+4\om(1+z)^3}}\,,
 \ee
which implies
 \be\label{wx0}
w_{\text{eff}}(0)=-{1\over 1+\om}\,.
 \ee

The DGP and LCDM models have the same number of parameters, with
$r_c$ substituting for $\Lambda$. (Neither model provides a
natural solution to the late-acceleration problem -- both
$\Lambda$ and $r_c$ must be fine-tuned to match observations.) The
DGP cosmology therefore gives a very useful framework for
comparing the LCDM general relativistic cosmology to a modified
gravity alternative. Does the DGP model pass the observational
tests? The next section considers this question.

\section*{3. OBSERVATIONAL CONSTRAINTS ON DGP}

The fundamental test of the background dynamics of a cosmological
model is the SNe magnitude-redshift test, based on the luminosity
distance,
 \bea
d_L &=& {(1+z) \over H_0\sqrt{|\Omega_K|}}\, {\cal F} \left(
\sqrt{|\Omega_K|}\int_0^z\, {dz' \over H(z')/H_0} \right),
\label{ld}
 \eea
where
 \bea
 {\cal F}(x) &\equiv & (x,\sin x, \sinh
x)~\text{for}~K=(0,1,-1)\,.
 \eea
The 68, 95 and 99\% likelihood contours from fits to the SNe
data~\cite{Riess:2004nr,Astier:2005qq} are shown in the DGP
parameter plane in Fig.~\ref{plane}.

The luminosity distance is determined purely by the modified
Friedman equation~(\ref{fz}), and there is a general relativistic
dark energy model that can exactly mimic the DGP luminosity
distance. This equivalent model has dark energy equation of state
equal to the DGP effective equation of state $w_{\text{eff}}$, as
given in Eq.~(\ref{wx}) for the flat case. Of course, there is no
physical justification for the equivalent general relativistic
model, whereas the DGP model has a clear physical motivation.
Nevertheless, tests based on the Hubble rate $H(z)$ cannot
distinguish the DGP from a GR dark energy model. Further tests are
needed to discriminate DGP from general relativistic models. The
CMB anisotropies and matter power spectrum provide in principle
suitable discriminatory tests. These tests require a detailed
understanding of the evolution of density perturbations in the DGP
model. Unfortunately the problem is highly complicated and not yet
solved, essentially since density perturbations are coupled to
perturbations in the 5D gravitational field~\cite{Koyama:2005kd}.

\begin{figure}
\begin{center}
\includegraphics[width=4.16cm]{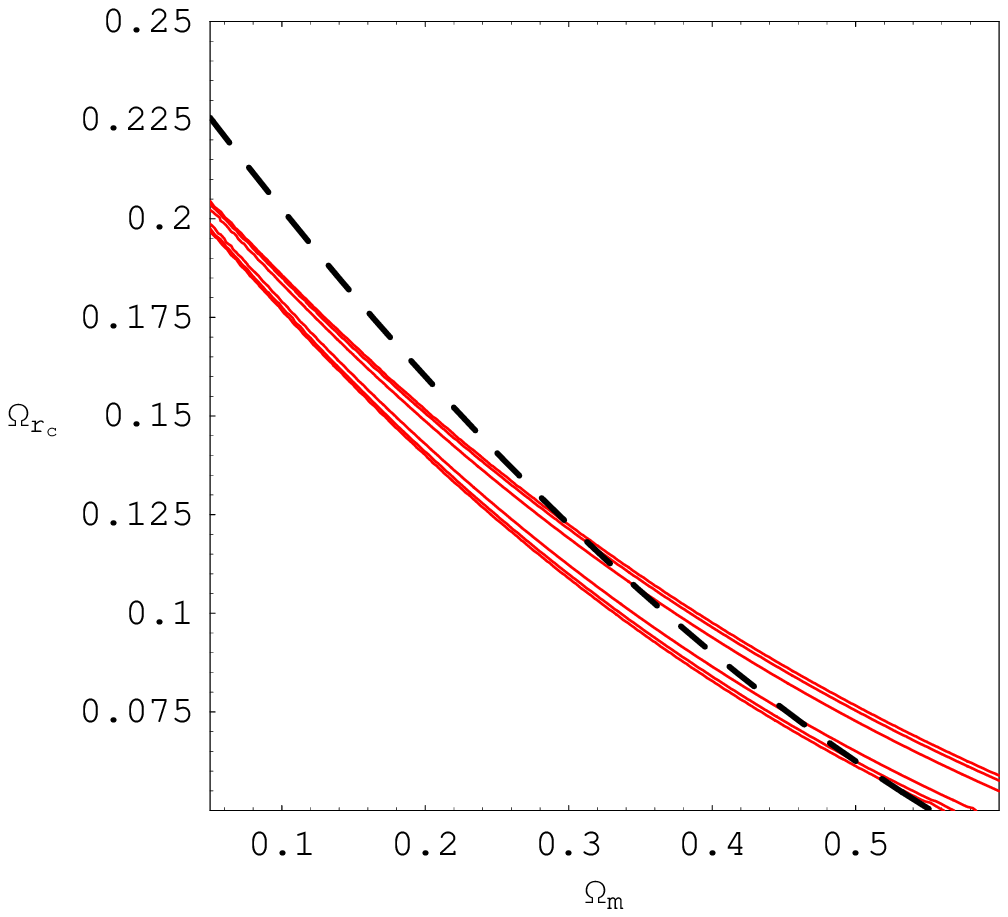}\quad
 \includegraphics[width=4.1cm]{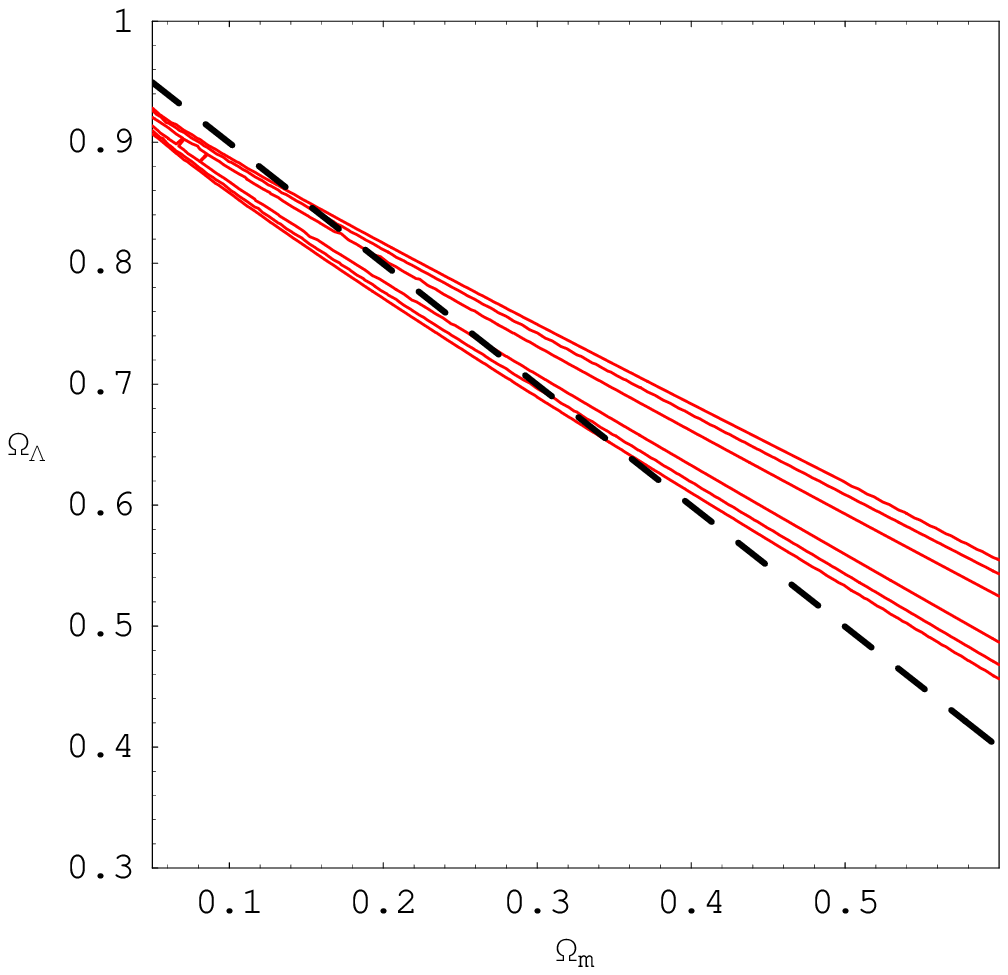}
\caption{Likelihood contours for the CMB shift parameter $S$ in
DGP (left) and LCDM (right). }\label{c}
\end{center}
\end{figure}

Previous results on the CMB anisotropies and structure formation
in DGP~\cite{Deffayet:2002sp} provide valuable strategies for
testing DGP and discriminating it from general relativistic
models. However, these results are based on neglecting the 5D
perturbations, which effectively imposes an inconsistency, leading
to a violation of the 4D Bianchi identity, as explained by Koyama
and Maartens~\cite{Koyama:2005kd}. On sub-Hubble scales, linear
density perturbations in the DGP model have been computed in a way
that satisfies the 5D equations~\cite{Koyama:2005kd}, and this
result confirms and generalizes the results of Lue and
Starkman~\cite{Lue:2005ya}. Further work is needed to quantify the
corrections to general relativistic perturbations on scales near
and above the Hubble radius.

Until the problem of cosmological perturbations in DGP is solved,
we must turn to other observations, independent of the SNe
redshifts, that rely on the background model and do not rely (or
rely only weakly) on the density perturbations.

\subsection*{CMB shift}

The CMB shift parameter
 \bea
S &=& \sqrt{\om}H_0\,{d_L(z_r) \over (1+z_r)} \,,
 \eea
relates the angular scale of the first acoustic peak to the
angular diameter distance to last scattering and the physical
scale of the sound horizon. It is effectively model-independent
and insensitive to perturbations, and may be used to constrain the
background dynamics of models, such as quintessence
models~\cite{Wang:2004py}. Wang and Mukherjee~\cite{Wang:2006ts}
have used the WMAP 3-year data~\cite{Spergel:2006hy} to derive
 \be
S=1.70\pm 0.03\,,
 \ee
with $z_r=1090$. The likelihood contours for $S$ in the DGP and
LCDM cases are shown in Fig.~\ref{c}. Note that the DGP contours
are more in the open-model region for $\om \lesssim 0.4$ than the
LCDM contours.

\begin{figure*}
\begin{center}
\includegraphics[width=8cm]{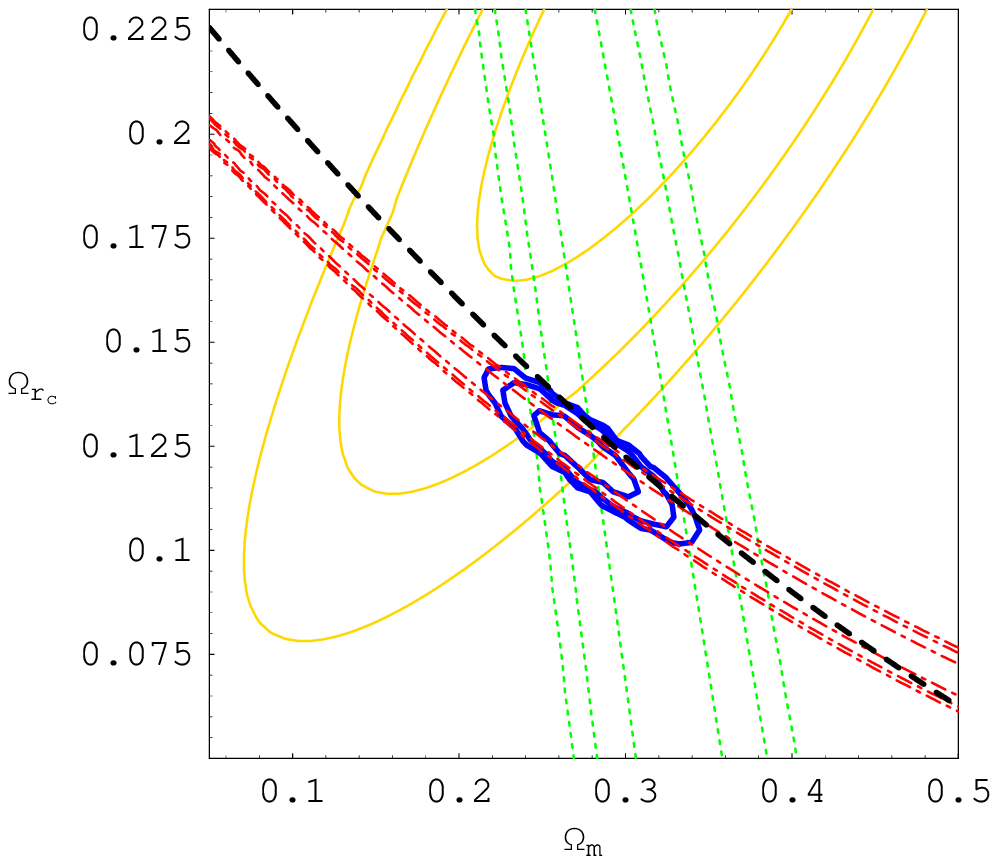}\quad
\includegraphics[width=8cm]{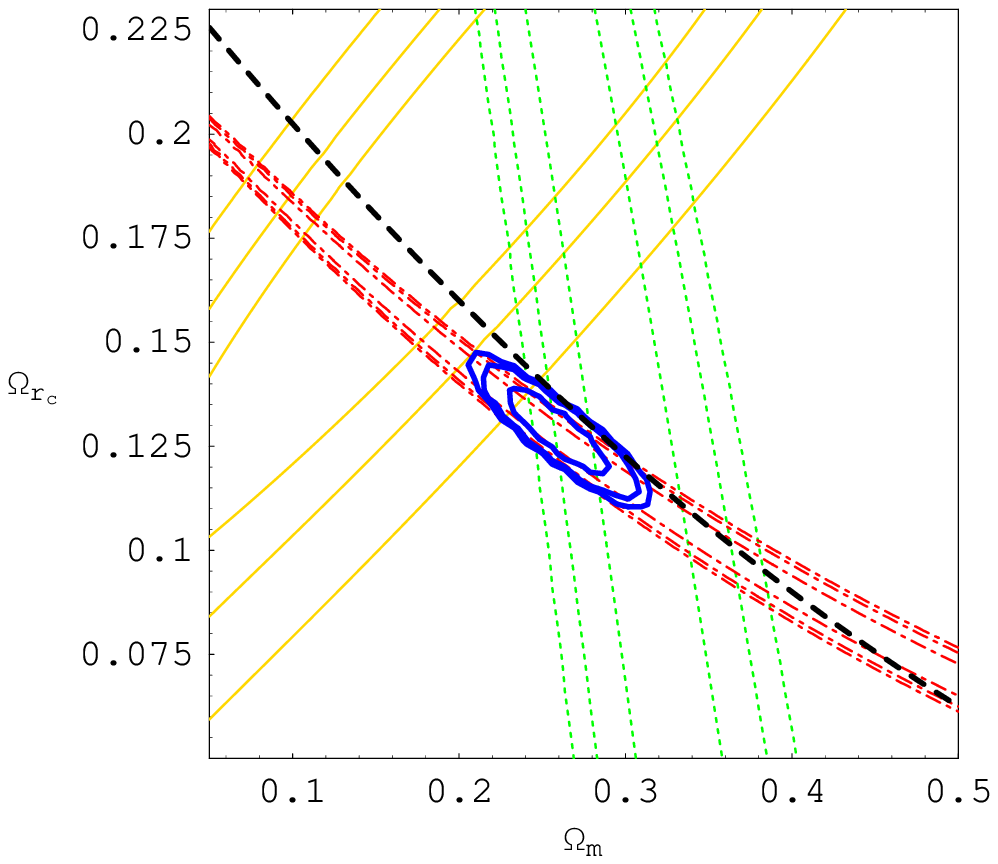}
\caption{Joint constraints [solid thick (blue)] on DGP models from
the SNe data [solid thin (yellow)], the BO measure $A$ [dotted
(green)] and the CMB shift parameter $S$ [dot-dashed (red)]. The
left plot uses SNe Gold data, the right plot uses SNLS data. The
thick dashed (black) line represents the flat models,
$\Omega_K=0$. }\label{sac}
\end{center}
\end{figure*}

\begin{figure*}
\begin{center}
\includegraphics[width=8cm]{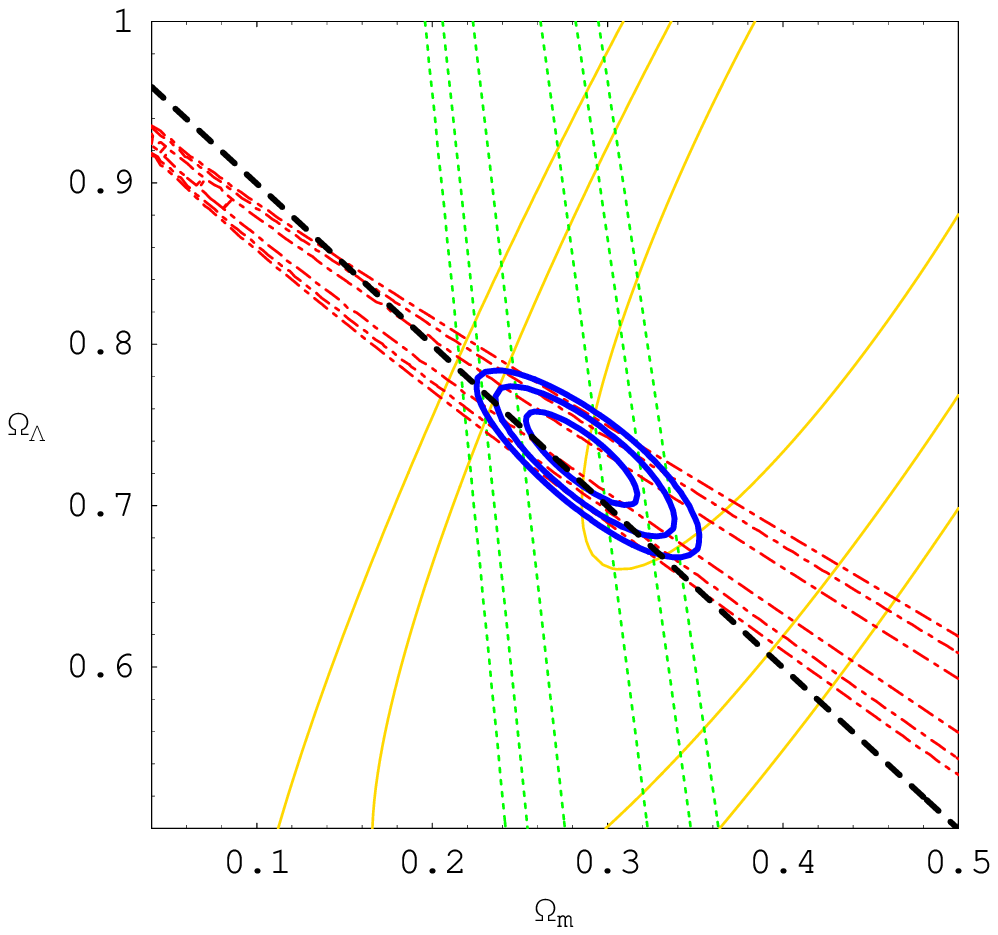}\quad
\includegraphics[width=8cm]{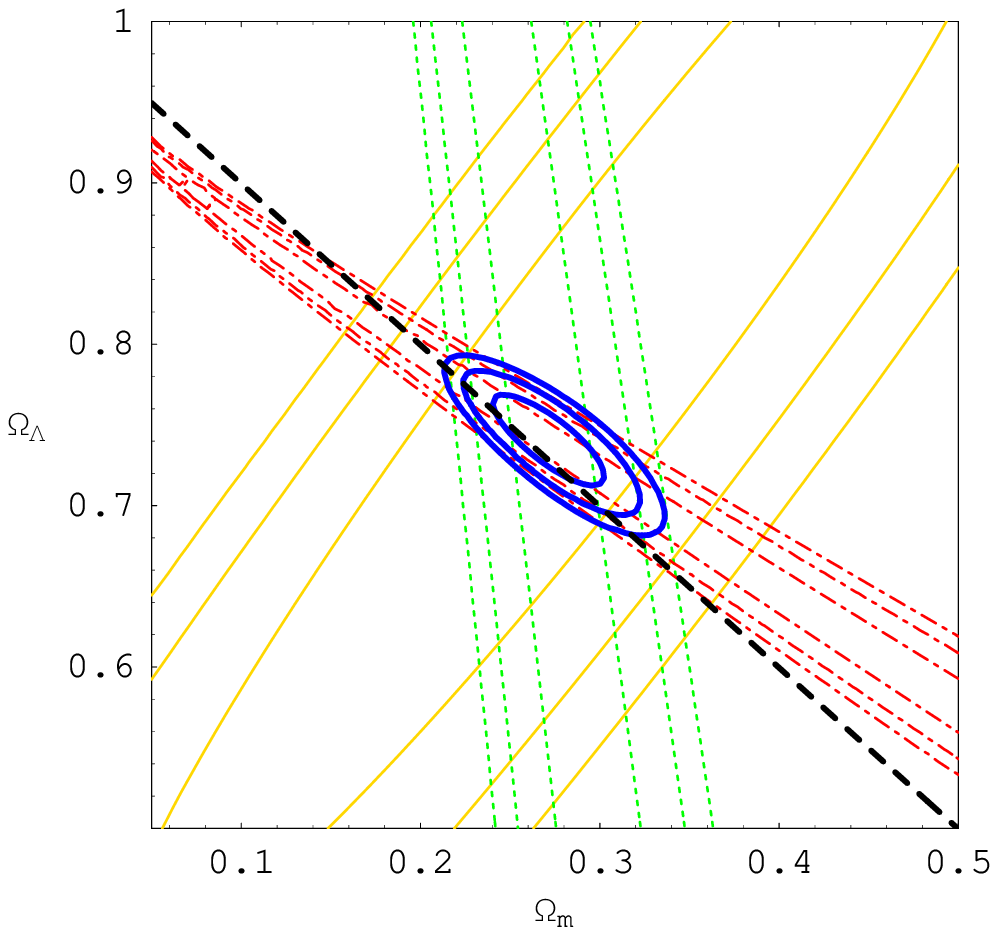}
\caption{As in Fig.~\ref{sac}, but for LCDM.}\label{sac_lcdm}
\end{center}
\end{figure*}

\subsection*{Baryon oscillations}

Fairbairn and Goobar~\cite{Fairbairn:2005ue} used the baryon
acoustic oscillation peak recently detected in the SDSS luminous
red galaxies (LRGs)~\cite{Eisenstein:2005su} as an independent
test of the DGP. The correlation function for SDSS LRGs shows a
peak at a scale $\sim 100 h^{-1}\,$Mpc, corresponding to the first
acoustic peak at recombination (determined by the sound horizon).
The observed scale effectively constrains the
quantity~\cite{Eisenstein:2005su}
 \bea
A &=& \sqrt{\om}\left[ {H_0^3\, d_L^2(z_1) \over H_1
z_1^2(1+z_1)^2} \right]^{1/3},
 \eea
and Eisenstein et al. find that~\cite{Eisenstein:2005su}
 \be
A= 0.469 \pm 0.017\,,\label{boa}
 \ee
where $z_1=0.35$ is the typical LRG redshift. (Following
Ref.~\cite{Fairbairn:2005ue}, we have suppressed a weak dependence
of $A$ on the spectral tilt.)

We should point out that there is a level of uncertainty in the
use of the BO measure $A$. The BO data are analyzed via a fiducial
LCDM model, with a single scaling relation used to make a best
fit, and a compression of the data to a constraint at a single
redshift~\cite{Eisenstein:2005su}. Within the LCDM class of
models, this approximation is accurate to within a few
percent~\cite{Eisenstein:2005su} (less than current errors in the
determination of $\om $). As pointed out by Dick et
al.~\cite{Dick:2006ev}, this single-redshift feature, which may be
robust for LCDM and constant-$w$ models, could introduce
significant errors for models with nonconstant $w$. The DGP has an
effective nonconstant $w$, although over the SDSS LRG redshift
range ($0.16\leq z\leq 0.47$) the variation is small. It is not
clear whether the errors introduced in fitting DGP models are at
the acceptable level of a few percent.  It is effectively
implicitly assumed that this is the case in
Refs.~\cite{Fairbairn:2005ue,Alam:2005pb,Guo:2006ce}. This may be
a reasonable assumption, since the DGP comoving luminosity
distance is close to LCDM in the redshift range for the
LRGs~\cite{Koyama:2005kd}. Ultimately, the BO data should be
re-analyzed without assuming a fiducial LCDM model and without
compression to a single redshift, so as to be a more reliable
constraint on alternatives to LCDM.

Another issue is the extent to which the BO constraint relies on
the LCDM matter power spectrum. The DGP corrections to the shape
of the matter power spectrum at last scattering on scales around
the first acoustic peak scale are small, although there may be
some scale-dependence in the corrections~\cite{kk}.

Fairbairn and Goobar showed that the joint constraints from SNe
(Legacy) data and the BO measure $A$ ruled out a flat DGP model at
3$\sigma$. Alam and Sahni~\cite{Alam:2005pb} used the Gold SNe
data with the BO measure $A$ and showed that the flat DGP model is
marginally allowed at 2$\sigma$. (See also
Refs.~\cite{Guo:2006ce}.) A conservative conclusion from these
results is that the SNe-BO constraints exclude flat DGP models at
2$\sigma$, unlike the case of LCDM models, where flat models are
within the joint 1$\sigma$ contour~\cite{Astier:2005qq}.

However, this negative conclusion is over-turned when we include
the CMB constraint from $S$.

\subsection*{Combined SNe-CMB-BO constraints}

The effect of the CMB shift $S$ is to pull the best-fit model down
from the closed region towards the flat curve. This is clearly
illustrated in Fig.~\ref{sac}: without the shift constraint, the
joint SNe-BO best-fit would be well above the flat model curve, as
found by Fairbairn and Goobar~\cite{Fairbairn:2005ue}.

The corresponding constraints for the LCDM model are shown in
Fig.~\ref{sac_lcdm}, and are consistent with the results of
Ichikawa and Takahashi~\cite{Ichikawa:2005nb}. For LCDM, the
SNe-BO constraints already allow a flat model at 1$\sigma$, and
the effect of the shift parameter is simply to narrow and rotate
the contours. It is clear from Figs.~\ref{sac} and \ref{sac_lcdm}
that there is some tension between the data and DGP, which is not
the case for LCDM.

\section*{4. CONCLUSIONS}

The combination of three independent observational constraints --
SNe, CMB shift parameter, baryon oscillations -- is a powerful way
to test the background dynamics of a cosmological model. It has
been used for quintessence models by Ichikawa and
Takahashi~\cite{Ichikawa:2005nb} and for an $f(R)$ modified
gravity model by Amarzguioui et al.~\cite{Amarzguioui:2005zq}. We
applied this triple joint constraint to the DGP case, using both
the Gold and the Legacy data, and showed that flat DGP models are
within the 2$\sigma$ joint contour. We note that the best-fit DGP
model is slightly open, as opposed to the slightly closed nature
of the best-fit LCDM model.

Although flat DGP and LCDM models are both within the 2$\sigma$
contour, it is clear from Figs.~\ref{sac} and \ref{sac_lcdm} that
LCDM fits the data more easily. A quantitative measure of this is
via the $\chi^2$ values for the best-fit models in the two cases,
as shown in Tables~I (Gold SNe data) and II (Legacy SNe data). The
tension between the data and the DGP is signalled by the rather
large value of $w_{\text{eff}}(0)$: using $\om\approx 0.26$ from
Tables~I and II, Eq.~(\ref{wx0}) gives
 \be
w_{\text{eff}}(0)\approx -0.8\,.
 \ee

\begin{center}
\begin{table}[!h]
\begin{tabular}{|c|c|c|c|c|} \hline
 & best-fit  & best-fit & best-fit &$\chi^2$
 \\ & acceleration & density & curvature& value \\
 & parameter & parameter &parameter &  \\\hline
 DGP & $\orc=0.125$ & $\om=0.270$ & $\Omega_K=+0.0278$ & 185.0\\
 ~LCDM~ & $~\Omega_\Lambda=0.730~$ &
 $~\om=0.285~$ & $~\Omega_K=-0.0150~$& $~177.8~$\\ \hline
\end{tabular}
\caption{Best-fit parameters from SNe (Gold)-CMB shift-BO
constraints, and $\chi^2$ values, for the DGP and LCDM models.}
\end{table}
\end{center}

\begin{center}
\begin{table}[!h]
\begin{tabular}{|c|c|c|c|c|} \hline
 & best-fit  & best-fit & best-fit &$\chi^2$
 \\ & acceleration & density & curvature& value \\
 & parameter & parameter &parameter &  \\\hline
 DGP & $~\orc=0.130~$ & $~\om=0.255~$ & $~\Omega_K=+0.0300~$ & $~128.8~$\\
 ~LCDM~ & $~\Omega_\Lambda=0.740~$ &
 $~\om=0.270~$ & $~\Omega_K=-0.0100~$& 113.6\\ \hline
\end{tabular}
\caption{As in Table~I, for the Legacy SNe data.}
\end{table}
\end{center}

For the Gold data, LCDM reduces the $\chi^2$ by 7, whereas the
Legacy data gives an even larger reduction of 15. LCDM provides a
significantly better fit to SNe-CMB shift-BO observations than
DGP, especially for the Legacy SNe data.

The $\chi^2$ per degree of freedom provides a comparison of the
goodness of fit across the two SNe data sets:
 \bea
\left(\chi^2/\text{dof} \right)_{\text{DGP}} &=& \left\{
\begin{array}{ll} 1.179 & \text{Gold} \\ 1.120 & \text{Legacy}
\end{array} \right. \nonumber \\
\left(\chi^2/\text{dof} \right)_{\text{LCDM}} &=& \left\{
\begin{array}{ll} 1.132 & \text{Gold} \\ 0.988 & \text{Legacy}
\end{array} \right. \nonumber
 \eea
The Legacy data leads to a better fit for both DGP and LCDM.

One qualification to our results is that we have not shown the
reliability of the BO measure $A$ as a clean test of the
background dynamics for DGP (this also applies to previous
applications of the BO constraint to DGP
models~\cite{Fairbairn:2005ue,Alam:2005pb,Guo:2006ce}). The BO
measure $A$ is tailored to the LCDM model, as explained above.
Although we may expect that only small corrections are involved,
it is not clear whether this introduces an artificial bias against
the DGP model. This is a nontrivial problem to resolve, and
further work is needed.

At a more fundamental level, the DGP needs to be tested against
the data on CMB anisotropies and the matter power spectrum, and
these tests provide also the means to discriminate DGP from the
equivalent general relativistic dark energy model. However, as
explained above, theses tests cannot yet be carried out, since the
analysis of the density perturbations in DGP has not yet been
sufficiently developed~\cite{Koyama:2005kd}. We should also note
that there is a ghost at the linear perturbative level in the
asymptotic de Sitter state of the self-accelerating DGP
cosmology~\cite{Gorbunov:2005zk}. Classically, this ghost is not a
problem, but it does render the quantum vacuum unstable, and we
effectively assume that an ultraviolet completion of the DGP model
may be found that cures the ghost instability in the quantum
vacuum.

\[ \]\[ \]
{\bf Acknowledgements:} The work of RM is supported by PPARC. We
thank Sidney Bludman, Rob Crittenden, Stephane Fay, Kazuya Koyama,
Ruth Lazkoz, Bob Nichol, Will Percival, Varun Sahni and Reza
Tavakol for useful discussions.

\end{document}